\documentclass[10pt, journal, onecolumn]{IEEEtran}  

\usepackage{graphics} 
\usepackage{epsfig} 
\usepackage[]{units} 
\usepackage{amsmath} 
\usepackage{amssymb}  
\usepackage{amsfonts} 
\usepackage{colonequals}
\usepackage[noend,ruled,noline]{algorithm2e}
\usepackage{bm}
\usepackage{siunitx}

\usepackage{booktabs}
\usepackage{color}
\usepackage{dblfloatfix}
\usepackage{url}
\usepackage{graphicx}
\usepackage[utf8]{inputenc}
\usepackage{fourier} 
\usepackage{array}
\usepackage{makecell}
\usepackage{cite}

\usepackage{mwe}

\usepackage{amsmath} 
\usepackage{graphicx}

\usepackage{float}

\usepackage{color}

\newcommand{\RR}{\mathbb{R}}
\newcommand{\CC}{\mathbb{C}}
\renewcommand{\vec}[1]{\mathbf{#1}}
\newcommand{\norm}[1]{\left\lVert#1\right\rVert}
\newcommand{\nth}[1]{#1^{\mathrm{th}}}

\makeatletter
\newcommand{\algstep}[1]{%
  \par
  \textit{#1\@addpunct{.}}\enspace\ignorespaces
}
\makeatother

\newcommand{\squeezeup}{\vspace{-2.5mm}}

\title{\LARGE \bf
Clinically Deployed Distributed Magnetic Resonance Imaging Reconstruction: Application to Pediatric Knee Imaging
}

\IEEEoverridecommandlockouts 
\author{
 \IEEEauthorblockN{Michael J.\ Anderson\IEEEauthorrefmark{1}\textsuperscript{$\diamond$}\thanks{\textsuperscript{$\diamond$}Equally contributed to this work.}\thanks{\copyright{} 2018 Intel Corporation.},
 Jonathan I.\ Tamir\IEEEauthorrefmark{3}\textsuperscript{$\diamond$}, Javier S.\ Turek\IEEEauthorrefmark{2}, 
 Marcus T.\ Alley\IEEEauthorrefmark{4}, Theodore L.\ Willke\IEEEauthorrefmark{2},\\
 Shreyas S.\ Vasanawala\IEEEauthorrefmark{4} and
 Michael Lustig\IEEEauthorrefmark{3}}\\
 \IEEEauthorblockA{\IEEEauthorrefmark{1}Intel Labs, Santa Clara, CA, United States}\\
 \IEEEauthorblockA{\IEEEauthorrefmark{2}Intel Labs, Hillsboro, OR, United States\\
 {Email: \{michael.j.anderson, javier.turek, ted.willke\}@intel.com}}\\
 \IEEEauthorblockA{\IEEEauthorrefmark{2}Electrical Engineering and Computer Sciences, University of California, Berkeley, CA, United States\\
 Email: \{mlustig,jtamir\}@berkeley.edu}\\
 \IEEEauthorblockA{\IEEEauthorrefmark{3}Radiology, Stanford University, CA, United States\\
 Email: \{mtalley,vasanawala\}@stanford.edu}
 \vspace{-10pt}
}

\begin{document}
\maketitle

\begin{abstract}
Magnetic resonance imaging is capable of producing volumetric images without ionizing radiation. Nonetheless, long acquisitions lead to prohibitively long exams. Compressed sensing (CS) can enable faster scanning via sub-sampling with reduced artifacts. However, CS requires significantly higher reconstruction computation, limiting current clinical applications to 2D/3D or limited-resolution dynamic imaging. Here we analyze the practical limitations to T2 Shuffling, a four-dimensional CS-based acquisition, which provides sharp 3D-isotropic-resolution and multi-contrast images in a single scan. Our improvements to the pipeline on a single machine provide a 3x overall reconstruction speedup, which allowed us to add algorithmic changes improving image quality. Using four machines, we achieved additional 2.1x improvement through distributed parallelization. Our solution reduced the reconstruction time in the hospital to 90 seconds on a 4-node cluster, enabling its use clinically. To understand the implications of scaling this application, we simulated running our reconstructions with a multiple scanner setup typical in hospitals.
\end{abstract}

\section{Introduction}
Magnetic resonance imaging (MRI) is a powerful medical imaging modality for visualizing tissue contrast,
and it is especially attractive for pediatric imaging as it does not involve ionizing radiation.
Nonetheless, imaging speed is a major limitation of MRI.
For example, MRI scans are orders of magnitude longer compared to computed tomography.
A typical MRI exam consists of several sequences, each providing different diagnostic image contrast.
As each sequence takes several minutes, the entire MRI exam can take 30 minutes to one hour.
This drives the cost up as only a small number of patients can be accommodated into the daily schedule.
In pediatrics, patients can become uncooperative over the course of the exam, necessitating the use of anesthesia \cite{bib:vasanawala2010}.  

Several advances to MRI acquisition have led to markedly faster scanning. Parallel Imaging (PI) uses multiple spatially sensitive receiver coils placed around the anatomy and exploits the redundancy in the receive channels to reduce the number of measurements required for reconstruction \cite{bib:sense,bib:grappa}. As the scan time is directly proportional to the number of acquired samples,
the speedup is proportional to the number of channels.
An additional avenue for faster scanning is through Compressed Sensing (CS) \cite{bib:lustig2007,bib:csmri}. In CS, signal sparsity is used to reduce the number of samples
required to perform reconstruction. Sparsity typically increases with the number of dimensions, making CS well-suited to MRI experiments involving image or signal dynamics.
Since PI and CS exploit fundamentally different properties
of the image and imaging system
to reduce sampling requirements, they can be combined
to realize order-of-magnitude reductions in scan time \cite{bib:hollingsworth2015,bib:yang2016}. We refer to their synergistic combination as PICS.

\begin{figure*}
\centering
\includegraphics[trim=10 5 10 15, clip, width=\textwidth]{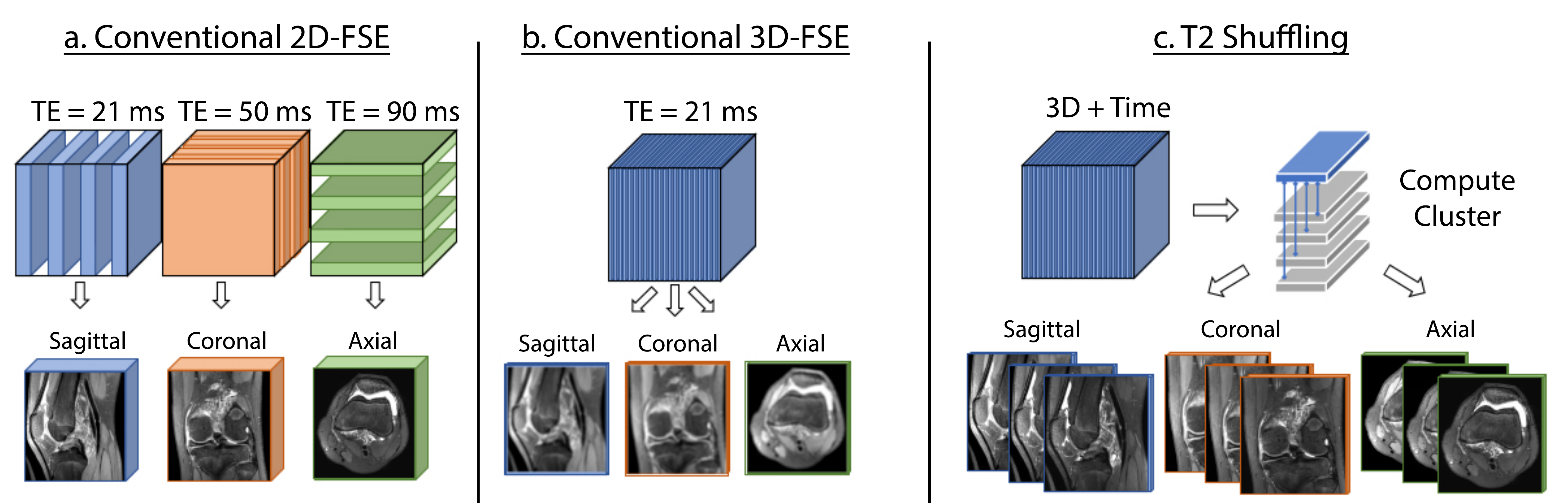}
\caption{(a) Conventional MRI exams separately acquire thick 2D slices at specific image contrasts and orientations. (b) 3D-FSE enables multi-planar reformatting,
but suffers from blurring. (c) T2 Shuffling improves image sharpness of 3D-FSE and simultaneously reconstructs multiple image contrasts.}
\label{fig:overview}
\squeezeup
\end{figure*}

T2 signal relaxation is an intrinsic MRI property of tissue, providing invaluable diagnostic image contrast. Volumetric T2-weighted imaging is typically obtained using a 3D fast spin-echo (3D-FSE) MRI acquisition \cite{bib:hennig1986,bib:mugler2014}, but the acquisition is lengthy and often results in image blurring due to the signal relaxation during the data collection \cite{bib:busse2008}.
T2 Shuffling \cite{bib:t2shmethod} is a particular PICS-based technique that mitigates these effects by modifying the
3D-FSE acquisition. From the data,
a time-series of 3D images are reconstructed, representing the temporal signal relaxation during the FSE acquisition. As a result,
the contrast in the time series of images transitions from proton-density weighting to heavy T2 weighting while not suffering from 3D-FSE blurring.
In particular, this technique is very promising for reducing the exam time in routine imaging applications, since the images can be reformatted and sliced into
arbitrary orientations as well as displayed at different levels of T2 contrast. This is shown in Figure~\ref{fig:overview}.
For example, T2 Shuffling has been applied to pediatric knee MRI \cite{bib:t2shclinical}, where small structures in the knee require oblique reformats at a high resolution. Other clinical applications are also leveraging 4D PICS-based approaches with promising initial results \cite{bib:zhang2014,bib:cheng2016,bib:feng2016}.

T2 Shuffling and other PICS approaches require iterative solvers for reconstruction \cite{bib:lustig2007}.
Therefore, the reduction in scan time is offset by the increased computational complexity.
As the acquisition and reconstruction methods grow in complexity, the computational requirements continue to grow, leading to reconstructions
that can take tens of minutes to several hours \cite{bib:wang2014,bib:lebel2014,bib:cheng2016}.
In order for these approaches to gain clinical traction, the image reconstruction pipeline should maintain a low latency,
so that the images are available at the scanner before the patient leaves the table. In our experience, conservative latency requirements for clinical integration are about two minutes.


Motivated by the need for fast PICS reconstructions, a number of general-purpose software packages have been built as a mechanism to translate the work
to the clinic \cite{bib:toolbox,bib:gai2013,bib:hansen2013,bib:knoll2014,bib:bartismrm,bib:ou2017,bib:lin2018}.
As a result, several applications of PICS have seen success in a clinical research setting \cite{bib:signapulse,bib:magnetomflash}.
Initial efforts focused on accelerating 3D reconstructions using CPUs and GPUs, achieving reconstruction times on the order of
one minute \cite{bib:stone2008,bib:kim2011,bib:murphy2012,bib:despres2017}.
Extensions to cloud-based processing \cite{bib:xue2015} have enabled reconstructions that take advantage of scalable compute.
More recently, the focus has shifted to 2D and 3D dynamic imaging \cite{bib:kowalik2015,bib:feng2016,bib:ting2017,bib:zhang2014}. The latter often involves significantly higher memory/computation requirements. 

In this work, we build a distributed PICS reconstruction using multiple high-performance computers and deploy it in a clinical setting. We specialize the reconstruction to T2 Shuffling, which despite its demonstrated benefits, has seen slow adoption in practice due to its prohibitively long reconstruction time \cite{bib:t2shclinical,bib:t2shmrvalue}.
We holistically analyze the entire reconstruction pipeline and the system requirements for successful deployment at the Lucile Packard Children's Hospital.
We describe the approaches we took to optimize the computational components, as well as analyze the scalability of the distributed computing.
Consequently, we reduce the end-to-end reconstruction time for T2 Shuffling from eight minutes on a single shared-memory system to under 90 seconds on a four-node distributed system.
As a result of the fast processing, we are able to improve image quality robustness by incorporating additional signal
processing steps into the reconstruction pipeline \cite{bib:roemer1990,bib:hansen2015,bib:sid} while maintaining
a suitable latency.
We present results for pediatric knee MRI, where the T2 Shuffling acquisition
and distributed reconstruction is used to replace multiple conventional 2D FSE scans,
enabling short exams that can be accommodated within the existing schedule \cite{bib:t2shmrvalue}.
Furthermore, we analyze the multiple-scanner setup in the hospital by simulating reconstruction events using real scanner schedule data. 
We present different configurations and sizes needed to support a target processing latency for our institution.


\section{MRI Reconstruction}
\label{sec:reconstruction}

The MRI reconstruction problem is often formulated as an unconstrained inverse problem using a prescribed forward model that describes the imaging experiment.

\subsection{Parallel Imaging and Compressed Sensing}\label{sec:pics}
A forward model for the MR sensing operator is that an image is multiplied by the
local spatial sensitivity profile of each coil, Fourier transformed, and discretely sampled \cite{bib:sense,bib:espirit}. Let $\vec x \in \CC^N$ represent a vectorized 3D
image with $N=N_xN_yN_z$ voxels.
The data formation process can be described for each coil as
\begin{align}
  \vec{y}_j = \vec{E}_j\vec x + \vec w_j = \vec{PFS}_j\vec x + \vec w_j, \quad j=1,\dots,C,
\label{eqn:coilmodel}
\end{align}
where the diagonal matrix $\vec{S}_j \in \CC^{N\times N}$ is the local spatial sensitivity profile of the $\nth{j}$ coil, $C$ is the number of coils, $\vec F\in\CC^{N\times N}$ is the 2D discrete Fourier transform, and the truncated diagonal matrix $\vec P = \{0,1\}^{P\times N}$ is a binary mask that is active at the locations of acquired k-space samples. The measurements from the $\nth{j}$ coil are given by $\vec y \in \CC^P$.
Complex white Gaussian noise $\vec w_j \in \CC^P$ is added to each measurement.
For convenience, we represent $\vec y$, $\vec w$, and $\vec P$ as $N$-dimensional, with $N-P$ zeros inserted in the locations that were not sampled.
The full forward model is thus described by $\vec y = \vec{E x} + \vec w$,
where $\vec y = \begin{bmatrix} \vec y_1^\top &\cdots & \vec y_C^\top \end{bmatrix}^\top$,
$\vec E = \begin{bmatrix} \vec E_1^\top & \cdots & \vec E_C^\top \end{bmatrix}^\top$, and
$\vec w = \begin{bmatrix} \vec w_1^\top &\cdots & \vec w_C^\top \end{bmatrix}^\top$.
It is possible to incorporate into the forward model prior knowledge about the system operating conditions.
For example, a ``soft-SENSE'' formulation \cite{bib:espirit} can be used to improve robustness to motion and to an under-prescribed field of view -- two realities that are seen in clinical practice \cite{bib:vasanawala2010}.
In the soft-SENSE formulation, $M$ sets of sensitivity maps are represented and $M$ image components are reconstructed.
Typically, the first image set captures the relevant image information, while the second (and more) image set captures data inconsistencies.
In this case, $\vec x = \begin{bmatrix}
(\vec x^{1})^\top & \cdots & (\vec x^M)^\top \end{bmatrix}^\top$, $\vec S = \begin{bmatrix}
\vec S^1 & \cdots & \vec S^M \end{bmatrix}$, $\vec x^m$ is the $\nth{m}$ image component,
and $\vec S^m$ is the $\nth{m}$ set of sensitivity maps.

The PICS reconstruction problem is often formulated as an unconstrained inverse problem \cite{bib:lasso,bib:csmri,bib:lustig2007},
\begin{align}
\arg\min_{\vec x}\frac{1}{2}\norm{\vec{y}-\vec{Ex}}_2^2 + Q(\vec x),
\label{eqn:pics}
\end{align}
where $Q(\cdot)$ is a regularization functional designed to promote sparsity.
The optimization problem \eqref{eqn:pics} can be iteratively solved with the Fast Iterative Soft-Thresholding Algorithm (FISTA) \cite{bib:beck2009}. Two main operations are sequentially applied in each iteration. In the first, the gradient of the data consistency term is computed and applied,
$
  \vec E^H\vec{E x} - \vec E^H \vec y,
$ 
where $\vec E^H$ represents the complex Hermitian transpose (adjoint) of the matrix $\vec E$.
In the second, the proximal operator of $Q$ is applied. When $Q$ is an $\ell_1$-like functional, this corresponds to soft thresholding.
The iteration complexity is dominated by these two components.

For 3D Cartesian imaging, it is possible to decouple the acquired data into independent 2D slices by performing an inverse Fourier transform in the fully sampled readout direction \cite{bib:murphy2012}.
Then, each slice is independently reconstructed and later rejoined into a 3D volume. Although this approach reduces the effective SNR of the reconstruction, it enables
parallel processing of each slice, say through multi-CPU or multi-GPU architectures.

\subsection{T2 Shuffling}\label{sec:t2sh}
In T2 Shuffling, the temporal signal dynamics due to T2 relaxation during a 3D-FSE acquisition are incorporated into the imaging formulation \cite{bib:t2shmethod}.
Specifically, it aims to recover
the T2 relaxation curve at each voxel, resulting in a set of 3D images. The time series of images are represented by
$\vec x = \begin{bmatrix}
\vec x_1^\top & \cdots & \vec x_T^\top \end{bmatrix}^\top \in \CC^{TMN}$, where $\vec x_i$ is the $\nth{i}$ timepoint (called an \emph{echo time}, TE) and
$T$ is the number of TEs (called the \emph{echo train length}).
Each TE image $\vec x_i$ is independently passed through the forward model and independently sampled, producing the measurement vector $\vec y_{ij} \in \CC^{N}$,
where $\vec y_{ij}$ represents the measurements from the $\nth{i}$ TE and $\nth{j}$ coil.

\begin{figure}
\centering
\includegraphics[width=0.95\columnwidth]{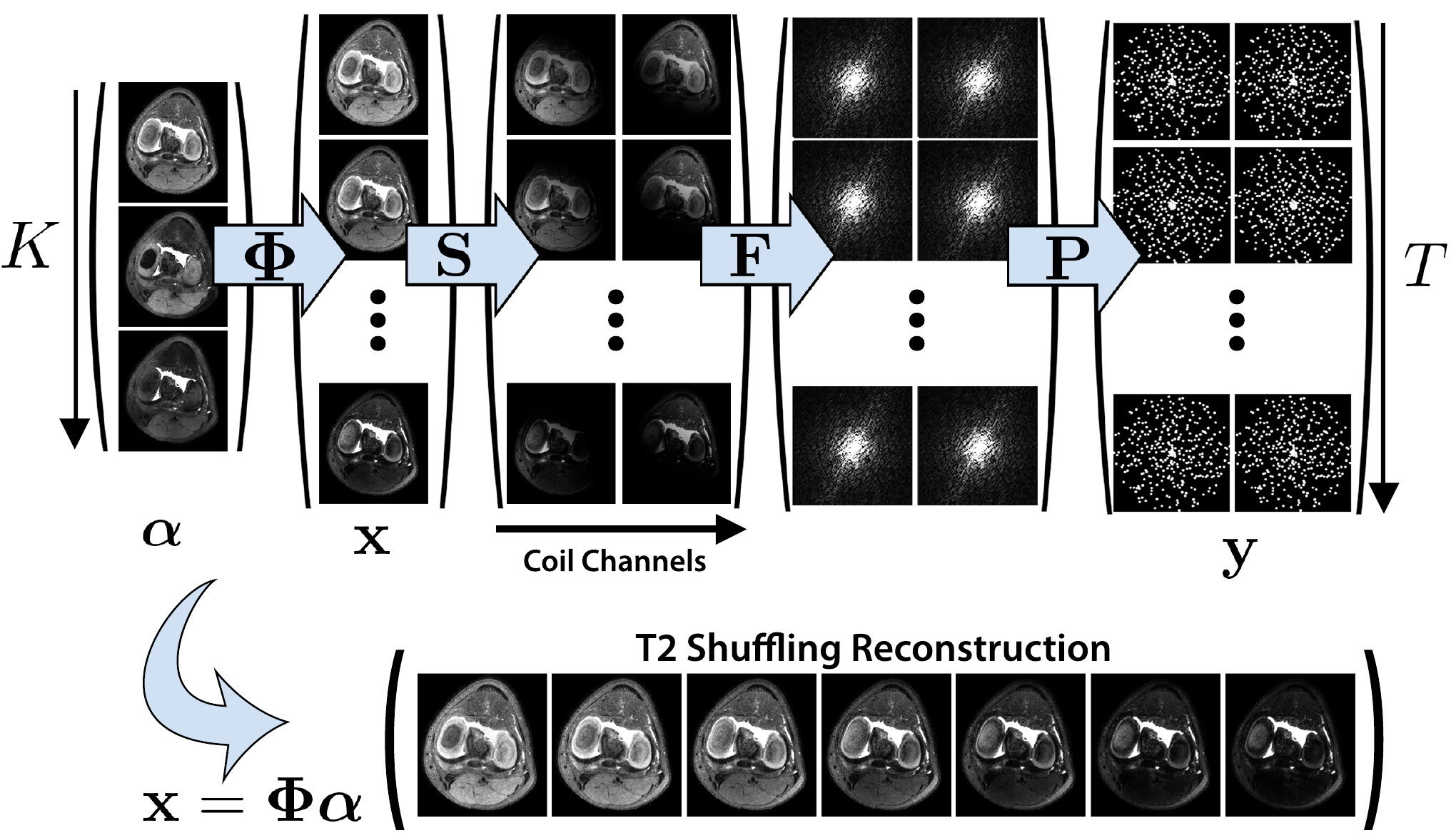}
\caption{Graphical depiction of the T2 Shuffling forward model for $K=3$ subspace coefficient images, $C=2$ coils, and $M=1$ set of sensitivity maps. The
reconstructed time series of images exhibit increasing T2 contrast.}
\squeezeup
\label{fig:t2shforwmodel}
\end{figure}

As the relaxation curves are smooth and correlated, the temporal signal dynamics are well approximated by a low-dimensional subspace \cite{bib:huang2012}.
Thus, the temporal signal evolution of the $\nth{n}$ voxel, $\vec x(n)$, is approximated as
\begin{align}
  \vec x(n) &\approx \bm{\Phi} \bm{\alpha}(n), \label{eqn:backproj}
\end{align}
where, $\bm \Phi \in \RR^{T\times K}$ is a matrix consisting of the basis vectors of the subspace of size $K$ and $\bm \alpha(n) \in \CC^{K}$ are the basis coefficients of $\vec x(n)$. The vector $\bm \alpha \in \CC^{KMN}$ represents the $K$ basis images consisting of $N$ voxels each.
Thus, $KMN$ images are jointly reconstructed. The forward model, accounting for the subspace representation, is shown in Figure~\ref{fig:t2shforwmodel}.

T2 Shuffling further exploits spatio-temporal correlations between neighboring voxels as \emph{locally lower rank} (LLR) \cite{bib:trzaskoismrm2011,bib:zhang2015}. This is accomplished by imposing a nuclear norm constraint \cite{bib:cai2010}
on spatio-temporal blocks extracted from the basis coefficient images $\bm \alpha$. More precisely, let
\begin{align}
  Q(\bm \alpha) &= \lambda \sum_{n} \norm{R_n(\bm \alpha)}_*,
  \label{eqn:llr}
\end{align}
where $R_n$ extracts a block of $B$ voxels centered around the $\nth{n}$ voxel
from each coefficient image $\bm \alpha_k$ and reshapes it into a $B \times K$ matrix, and $\lambda$ is the regularization parameter. The nuclear norm proximal map for each
patch is solved via singular value soft thresholding \cite{bib:cai2010}, and can be computed independently for each matrix. For practical reasons, the coefficient images are randomly
shifted each iteration and non-overlapping blocks are extracted; this stochastic approach reduces blocking artifacts due to boundary conditions without computing fully overlapping blocks.

Putting \eqref{eqn:pics}, \eqref{eqn:backproj}, and \eqref{eqn:llr} together, The T2 Shuffling reconstruction problem is formulated as
\begin{align}
  \arg\min_{\bm \alpha} \frac{1}{2}\norm{\vec y - \vec{P} \vec{F S}\bm\Phi\bm\alpha}_2^2 + \lambda\sum_{n}\norm{R_n(\bm \alpha)}_*,
  \label{eqn:t2shpics}
\end{align}
where the operators are defined in a manner consistent with \eqref{eqn:coilmodel}.
After solving \eqref{eqn:t2shpics}, a spatial wavelet soft-threshold is applied to each basis coefficient image for additional spatial denoising.
The time-series of images are then recovered through \eqref{eqn:backproj}.
The T2 Shuffling reconstruction has also been extended to include partial Fourier sampling \cite{bib:santamarta2004,bib:t2shpf}, in which approximate conjugate symmetry
is used to reduce the number of acquired measurements. Following the reconstruction, a Homodyne filter \cite{bib:noll1991} that accounts for low-frequency phase is applied to $\vec x$.

\subsection{Clinical Integration}
To run T2 Shuffling in a clinical workflow, it is necessary to integrate \eqref{eqn:t2shpics} into a full reconstruction pipeline. This includes processing the
vendor-formatted raw data, computing the basis based on the scan parameters, estimating coil sensitivity maps, applying gradient non-linearity
corrections, converting the result to DICOM images, and transferring the images to the Picture Archiving and Communication System (PACS).
The full pipeline must maintain a low latency so that the images are available at the scanner console before the patient leaves the table.
The pre-processing and post-processing overhead is a non-negligible
portion of the overall latency; these aspects are often not accounted for when comparing algorithm runtimes.

\section{Baseline Implementation}
\label{sec:system_implementation}

In what follows, we describe the main computational steps of the baseline implementation of T2 Shuffling, with which we later compare our optimized solution in Section \ref{sec:results}.
The baseline implementation used the Berkeley Advanced Reconstruction Toolbox (BART) \cite{bib:bartismrm,bib:bart}, a software programming library intended for high-dimensional PICS applications.
We describe the pipeline in three sections: (A) a pre-processing stage that computes all the information needed for the iterative slice-by-slice reconstruction; (B) the iterative reconstruction stage that computes the basis coefficient images; and (C) post-processing stage that processes the reconstruction and feeds the resulting DICOM images back to the scanner console and to PACS.

\subsection{Pre-processing}
\algstep{Conversion to BART Format}
As a first step the acquired k-space data, consisting of an array of interleaved single precision
complex floats, are converted from the vendor-proprietary format to the BART format.
The measurements are stored as a sparse array consisting of 3D spatial frequency index, TE index, and coil channel index.

\algstep{Re-sort and Average}
From the sparse array, the measurements corresponding to the central 24-by-24-by-24 window are extracted from each channel,
averaged across the TEs, and reformatted as a 4D array. We refer to this low-resolution k-space as the calibration data.

\algstep{Coil Compression}
From the calibration data, a coil compression matrix is computed and applied
to the raw data \cite{bib:zhang2013}.
The purpose of coil compression
is to reduce the computational cost of large receive arrays in the reconstruction by linearly combining the data from multiple coils into a smaller number of virtual coils. This process is done with little to no loss of information. 

\algstep{Scale Factor}
As the receive gain changes from scan to scan, a scaling normalization factor is computed and applied. The calibration data
are inverse Fourier transformed and root-sum-of-square coil-combined to form a low-resolution image. The scaling is computed as the $\nth{95}$ percentile of the coil-combined image.

\algstep{ESPIRiT Maps}
From the low-res data, soft-SENSE coil sensitivity profiles are estimated 
using ESPIRiT \cite{bib:espirit}, an auto-calibrating PI method.

\algstep{Create Basis}
The RF refocusing flip angle schedule is used to compute a temporal basis that matches the particular acquisition \cite{bib:t2shmethod}. The basis is computed by simulating an ensemble of signal evolution curves with the extended phase graph algorithm \cite{bib:weigel2015},
performing a singular value decomposition (SVD) on the matrix of simulated signals, and retaining the $K$ largest singular vectors.

\algstep{Re-Sort and Project Data}
Using the basis, the sparse-array data are re-sorted and projected onto the basis, yielding a 5D data set of size $\begin{bmatrix}N_x&N_y&N_z&C&K\end{bmatrix}$, where $N=N_x N_y N_z$ is the 3D image matrix size, $C$ is the number of coil-compressed channels, and $K$ is the subspace size.

\algstep{Compute Mask} As a last step, the $T$ spatial sampling patterns are computed from the sparse array and stored as a zero-filled, one-zero mask for each time point.

\subsection{PICS Reconstruction}
Prior to the iterative reconstruction, a 1D unitary inverse Fourier transform is applied in the readout ($N_x$) direction of the 5D
projected k-space data to decouple the readout slices.
The reconstruction is then performed independently on each slice, and multiple slices are reconstructed in parallel based on the number of available CPU cores.
\algstep{Linear Recon}
As the baseline implementation resulted in latency times exceeding 5 minutes, a fast and low-resolution linear reconstruction (with Tikhonov regularization in place of LLR regularization)
is first solved to provide image quality feedback to the technologist before releasing the patient.
\algstep{Detailed Recon}
The full iterative reconstruction is performed to solve the optimization problem in \eqref{eqn:t2shpics} using FISTA.
Following the iterative reconstruction, the slices are re-joined into a 5D volume for post-processing.

\subsection{Post-processing}
\algstep{Process TE}
In the post-processing step, the 5D reconstructed volume is denoised with a single iteration of 3D wavelet soft-thresholding to reduce noise amplification.
After, the reconstructed images are multiplied by the sensitivity maps and Fourier transformed to synthesize fully reconstructed k-space data.
The data are back-projected with the basis at each desired echo time point, and Homodyne filtering is applied.
The result is corrected for gradient non-linearities, converted into the DICOM format, and transferred back to the scanner and to the hospital PACS.
In the baseline implementation, three time points are sequentially processed, corresponding to a
proton-density weighted image (TE = 21 ms), intermediate weighted image (TE = 50 ms), and T2 weighted image (TE = 90 ms).

\section{Optimizations}
\label{sec:optimization}

In this section, we describe strategies for speeding up the baseline implementation of T2 shuffling, as well as improvements to image quality which are enabled by the speedups. We identify and parallelize several serial bottlenecks in the BART toolbox (e.g. FFT phase modulation, normalization, and basis generation). We also exploit inter-operator fusion to improve cache behavior and eliminate costly transpose operations. Some computations are also completely removed from the pipeline using pre-computation and/or code refactoring. We also distribute the computation across multiple machines by exploiting parallelism across image slices in the PICS portion, and exploiting parallelism across time points during post-processing. This improved performance allows us to add additional signal processing steps to the pipeline, which improves image quality. 

\subsection{Improving Multi-core Parallelism}
\label{subsec:optimization_multicore}
\algstep{FFTmod}
Before running FFT, the image data must be un-centered, either through a circular shift
or through a phase modulation, multiplying each data element to add a shifting phase that is relative to its position in the multi-dimensional array.
While this operation is embarrassingly parallel, its parallelization requires several modulo divisions for the computation of each element's index.
Fortunately, when the dimensions of the array are even, for consecutive elements the phase alternates between $+1$ and $-1$, or $i$ and $-i$. 
As an optimization, we pre-select even-sized dimensions for images and optimize the phase modulation by parallelizing over blocks of consecutive elements. 
For each block, we compute the position and phase for the initial element, and alternate the phase for the remaining elements.

\algstep{Sort}
The baseline code in BART used Quicksort in the process of estimating scaling normalization factors. We replace the call to Quicksort with GNU parallel sort. 

\algstep{Basis Generation}
As the basis is computed by simulating an ensemble of signal time courses, it is embarrassingly parallel across the different signal simulations. We take advantage of this
fact during pre-processing to simulate each signal evolution on a single CPU core, and parallelize the simulations across all the cores of the machine.
In each simulation, the signal value is propagated through the time points by repeatedly applying matrix operations that represent the acquisition dynamics \cite{bib:weigel2015}.
We further reduce the simulation time by pre-computing common matrix terms that do not change during the simulation.

\algstep{Multi-dimensional Operations}
We also looked at BART itself to find opportunities for performance improvement through more efficient uses of parallelism. From the baseline BART implementation, we modified the parallel \texttt{md\_nary} operation, which applies an arbitrary function to many elements in a multi-dimensional array, to be a flat parallel loop instead of a nested for loop. We also added a configuration capability for the loop chunk size and tuned it for coil compression and data re-sorting.

\algstep{Process TE}
We reduce the post-processing time by operating directly on the coil-compressed data, reducing coil-by-coil processing (e.g. Homodyne filter, FFT). Additionally, we process the three time points in parallel, allocating one third of the machine’s cores to each TE.

\subsection{Fusing Forward and Adjoint Linear Operators}
\label{subsec:optimization_fusion}
Due to the nature of FISTA, the forward and adjoint operators are often applied in sequence (i.e. $\vec{E}^H \vec{E} \bm\alpha$).
We can identify several optimizations that are not otherwise possible when applying these operators naively. 

The first optimization which takes advantage of this is already included in the baseline \cite{bib:t2shmethod}. Noticing that $\bm \Phi$ operates across time 
and that $\vec F$ and $\vec S$ operate across space, the operators commute, i.e.
$
  \vec{FS}\bm \Phi = \bm \Phi \vec{FS}.
  \label{eqn:commute}
$
Thus, the forward and adjoint operation is equivalent to
$
  \vec S^H \vec F^H \bm \Phi^\top \vec P \bm \Phi \vec F \vec S.
  \label{eqn:normaleq}
$
Defining $\bm{\Psi}(n) \colonequals \bm{\Phi}^\top \vec{P}(n) \bm{\Phi} \in \CC^{K\times K}$ as the \emph{space-time kernel} matrix associated with the $\nth{n}$ spatial frequency,
we pre-compute and store $\bm{\Psi} = \{\bm{\Psi}(n)\}_{n=1}^N$ as a $K \times K \times N$ dense array. In this work, we further exploit the property that each $\bm{\Psi}(n)$ is real-valued and symmetric. We modify the storage to include only the real part of the matrix which reduces memory loads of this matrix by a factor of two. We also take advantage of the symmetry by loading only half of the matrix, cutting down on memory accesses by an additional factor of two. 

\begin{figure}
\centering
\includegraphics[trim=70 0 10 0, clip, width=0.6\columnwidth]{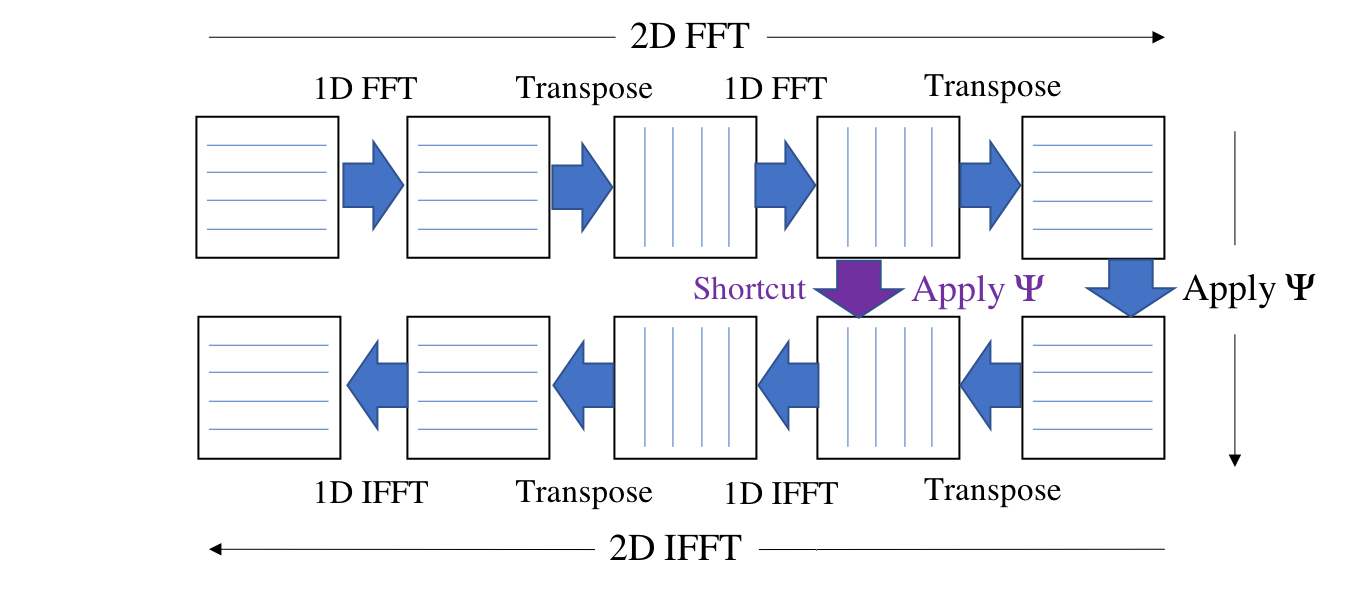}
\caption{Graphical depiction of the fused forward and adjoint T2 Shuffling operators. We avoid two transpose operations by leveraging the fact that the $\Psi$ operator is agnostic to the image data ordering.}
\squeezeup
\label{fig:fusion}
\end{figure}

We also take advantage of the structure of the joint application of the forward 2D Fourier transform followed by inverse 2D Fourier transform (i.e. $\vec{F}^H \bm{\Psi} \vec{F}$). This operation is shown visually in Figure \ref{fig:fusion}. The 2D Fourier transforms are applied typically by doing a set of 1D FFTs on the rows, followed by a matrix transposition, then a set of 1D FFTs again on the rows, followed by a matrix transposition to return the data to its original format. The lines in the figure correspond to the matrix ordering after each operation (ie row or column). Since the application of $\bm{\Psi}$ is an independent computation across voxels, it is agnostic to the ordering. Therefore, we can leave the argument in transposed form in between the forward and inverse Fourier transform computations, thus avoiding two of four transpose operations. This optimization is shown as the shortcut path in the figure. 

\begin{algorithm}
\LinesNumbered
\DontPrintSemicolon
 \KwData{Coefficient image data $\bm{\alpha}$, SENSE maps $\vec{S}$, coefficient space-time-kernel matrix $\bm{\Psi}$, temporary image storage $\vec t_{1}$ and $\vec t_{2}$}
 \KwResult{Coefficient image data transformed by the normal equations,
 $\bm{\alpha}^{\mathrm{out}} = \vec{S}^H \vec{F}^H \bm{\Psi} \vec{F} \vec{S}\bm\alpha$}
 \For{each sensitivity map $\vec{S}_i$}{
   \For{each panel $P$ in parallel} {
     \For{each coefficient image $\bm\alpha_j$} {
       \For{each row $r$ in $P$}{
            Set $t_{j,1}$ to $0$ Row $r$\;
            Apply $\vec{S}_{i,0}$ to $\bm\alpha^{0}_{j}$ and add to $\vec t_{j,1}$ Row $r$\;            
            Apply $\vec{S}_{i,1}$ to $\bm\alpha^{1}_{j}$ and add to $\vec t_{j,1}$ Row $r$\;            
            Apply $\vec{F}$ to $\vec t_{j,1}$ Row $r$\;
           }
       }
       Transpose $\vec t_{j,1}$ panel $P$ and store in $\vec t_{j,2}$\;
     }
   \For{each panel $P$ in parallel} {              
     \For{each row $r$ in $P$}{
       \For{each coefficient image $t_{j,2}$} {
           Apply $\bm{\Psi}$ to $\vec t_{j,2}$ Row $r$\;      
           Apply $\vec{F}^H$ to $\vec t_{j,2}$ Row $r$\;
         }
       }
       \For{each coefficient image $\vec t_{j,2}$} {
         Transpose $\vec t_{j,2}$ panel $P$ and store in $\vec t_{j,1}$\;
       }
     }
     \For{each panel $P$ in parallel} {             
       \For{each row $r$ in $P$}{
         \For{each coefficient image $\vec t_{j,1}$} {
         Apply $\vec{F}^H$ to $\vec t_{j,1}$ Row $r$\;
         \lIf{$i$ is $0$}{
         	Set $\bm\alpha^{\mathrm{out}}_j = 0$
         }         
	 Apply $\vec{S}^{H}_{i,0}$ to $\vec t_{j,1}$ and add to $\bm\alpha^{\mathrm{out}}_j$ Row $r$\;         
	 Apply $\vec{S}^{H}_{i,1}$ to $\vec t_{j,1}$ and add to $\bm\alpha^{\mathrm{out}}_j$ Row $r$\;         
       }
      }
    } 
  }
\caption{Parallel Fused Forward and Adjoint Linear Operator}
\label{alg:operator}
\end{algorithm}

We also fuse and reorder loops in the forward and adjoint operators in order to increase reuse in cache and reduce traffic to and from memory. This is shown in Algorithm \ref{alg:operator}. We improve reuse in cache by fusing the sensitivity map application of each image row to the FFT computations in both the forward (lines $6-8$) and adjoint (lines $20-23$) phases. We also move the loop over sensitivity maps to the outermost position (line $1$). This means that instead of computing $CK$ intermediate vectors, we take one sensitivity map and compute only $K$ intermediate vectors at a time, reducing this portion of the working set by a factor of $C$ which allows for better use of cache.

Finally, we parallelize the new optimized forward and adjoint operators across image panels (lines $2$, $10$, and $17$), which are groups of roughly $10-20$ consecutive rows assigned to a single thread. This creates two levels of coarse-grained parallelism: (1) the volume split  across slices, and (2) across panels within a slice. The second level of parallelism wasn't initially necessary due to the amount of parallelism available across slices, however it became useful for two reasons. First, by increasing the number of cores applied to each problem, we increase the likelihood that intermediate results will fit in on-chip memory. The benefit of this effect is shown in more detail in Section \ref{sec:results}. Secondly, this allows us to scale to more cores than the number of slices, which in our experiments is 288. 

\subsection{Removing Unnecessary and/or Redundant Computations}
\label{subsec:optimization_computation}
\algstep{Fast Linear Reconstruction}
The original pipeline included a fast linear reconstruction to provide quick feedback to the technologist.
As the overall latency of the detailed reconstruction became nearer to that of the linear reconstruction due to our optimizations, we removed the linear reconstruction altogether.

\algstep{Basis and Space-time Kernel Caching}
The basis operator $\bm{\Phi}$ is determined by the target scan anatomy (e.g. knee), the refocusing RF pulse flip angle schedule, and the associated echo spacing between the pulses during the data acquisition. As it does not depend on the particular patient scan, it can be stored and re-used for future scans that have the same acquisition
parameters. To accomplish this, we calculate a SHA-1 hash of the list of flip angles, echo spacing, and target anatomy.
Before computing the basis, we search for the hash in a look-up table to determine if the basis has been previously computed. If it is not found, we compute and store the matrix in the look-up table.
In a similar fashion, we pre-compute and store the space-time kernel $\bm{\Psi}$ in a separate look-up table.
The SHA-1 hash of $\bm{\Psi}$ consists of the SHA-1 of the basis and of the sampling pattern. Since the basis and the space-time kernel are 7.1 KB and 7.7 MB, respectively,
we found the load times to be negligible.

\algstep{Single-pass Locally Lower Rank Regularizer (LLR)}
Recall from Section~\ref{sec:reconstruction} that T2 Shuffling uses LLR regularization to exploit spatio-temporal correlation within the basis coefficient images.
The LLR operates on patches of the coefficient images, which are size $b \times b$ and non-overlapping in our implementation.
In each solver iteration, the coefficient images are first circular-shifted randomly in both dimensions by a small amount less than the patch size. Then, the patches are extracted from the $K$ coefficient images and rearranged into a 2D matrix per patch with one column per coefficient image and $b \times b$ rows. Common factors are extracted across coefficient images using the SVD, and any singular values below a threshold are set to zero before regenerating the coefficient images back from the factors. 


We modified this procedure work in a single pass through the coefficient images in order to reduce memory traffic and unnecessary copies.
Each thread allocates a temporary buffer of size $b^2 \times K$ ahead of time. Then we pass over the image, extracting data according
to the parameters of the circular shift, without explicitly performing the shift. We perform the SVD threshold operation in the thread-local buffer,
and scatter the data back to the original coefficient image when all the work for that patch is complete.

\subsection{Image Quality Improvements}
\label{subsec:optimization_improved}

The reduction in processing time enabled additional signal processing steps to be added with the aim of improving robustness across a diverse set of pediatric patients.
Specifically, we used Auto-ESPIRiT \cite{bib:sid}, a parameter-free ESPIRiT implementation in which the internal parameters are chosen based on Stein's Unbiased Risk Estimate
(SURE). Auto-ESPIRiT produces more consistent sensitivity maps in cases with a high degree of data inconsistency, e.g. due to motion or noise. To use Auto-ESPIRiT, an
accurate noise estimate is needed. To this end, we incorporated channel noise pre-whitening \cite{bib:roemer1990,bib:hansen2015} as a step in pre-processing. As the SURE-based parameter estimate can be computed for each slice independently, it is performed immediately before the PICS reconstruction.

\subsection{Distributed Implementation}

\label{subsec:optimization_distributed}

\begin{table}
  \caption{Sequence and reconstruction parameters.}
\label{tab:params}
\centering
\small
\begin{tabular}{l | c | c}
  \toprule
  Description & Symbol & Size \\
  \hline
  Spatial matrix size & $N_x\times N_y\times N_z$ & 288$\times$260$\times$240 \\
  Receive channels & $C_{\mathrm{total}}$ & 16 \\
  Coil compressed channels & $C$ & 7 \\
  Echo train length & $T$ & 80 \\
  Soft-SENSE maps & $M$ & 2 \\
  Basis coefficient images & $K$ & 4 \\
  LLR patch size & $b\times b$ & 12$\times$12 \\
  LLR regularization parameter & $\lambda$ & 0.0125 \\
  FISTA iterations & $N_{\mathrm{itr}}$ & 250 \\
  ESPIRiT calibration size & $r_x\times r_y\times r_z$ & 288$\times$24$\times$24 \\
  \bottomrule
\end{tabular}
\squeezeup
\end{table}

\begin{figure}
\centering
\includegraphics[width=0.6\columnwidth]{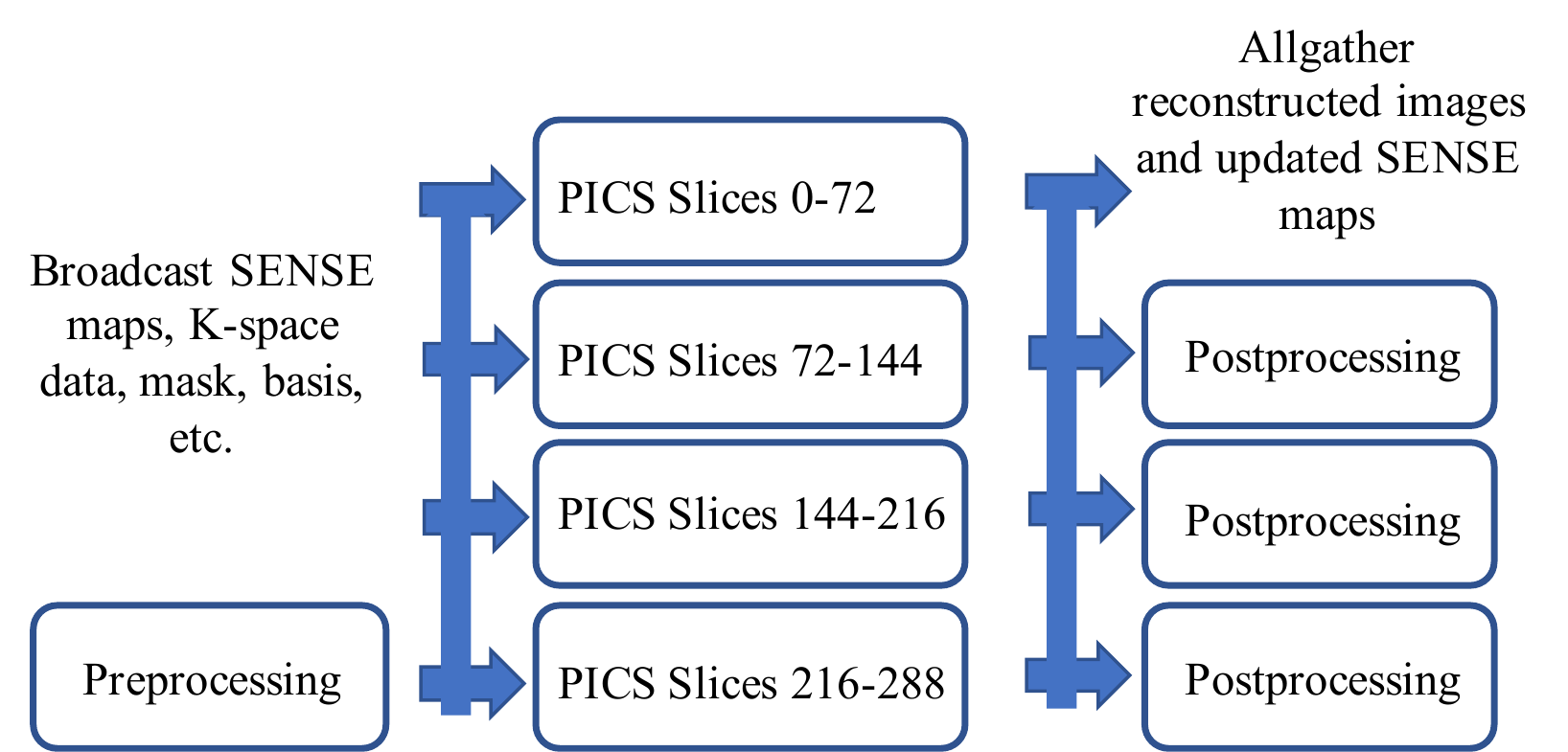}
\caption{Distributed Implementation including Broadcast and Allreduce MPI primitives.}
\vspace{-6mm}
\label{fig:distributed}
\end{figure}

Figure \ref{fig:distributed} depicts our implementation of the distributed reconstruction pipeline. We distribute The PICS portion and the post-processing over multiple machines in the cluster, while the pre-processing happens on a single machine.

The PICS portion is divided across slices with approximately equal numbers of slices sent to all machines. In order to do this, we must send the pre-processed SENSE maps, k-space data, sampling pattern, basis matrix, and other files to each machine. To do this, the files are loaded into Python from \texttt{/dev/shm}, communicated using the \texttt{mpi4py} library, and written to \texttt{/dev/shm} on the destination machines. We use \texttt{/dev/shm} extensively as the system uses the filesystem to store intermediate results from multiple calls to BART. The Broadcast primitive is used for all data, but it is is not always needed in cases where a scatter would suffice, such as for the k-space data and sensitivity maps. However we use broadcast for convenience, and to avoid local volume slicing overheads. Once the data is distributed, the PICS computation proceeds in a hybrid fashion with MPI parallelism across slices and OpenMP parallelism within each slice.

The post-processing portion is distributed over only three machines using \texttt{mpi4py}, as there are currently three time points being post-processed. In order to do this, the machines each need all the k-space data and the updated SENSE maps to proceed with post-processing generating the final images. We use \texttt{mpi4py} again to achieve this data transfer. We use the Allgather primitive to collect the data to these three machines. It is not strictly necessary to use Allgather, as only three machines need all the data. However we do this for convenience and find that the cost of this primitive is relatively small in comparison to the entire reconstruction runtime for the small cluster configurations we are considering.

Pre-processing is left to run on a single machine so as to not over-complicate the implementation. Our plan is to expand this distributed framework to other reconstruction methods with a similar parallelism profile. Such a framework would be complicated by excessive use of distributed primitives within areas such as pre-processing that we'd like to make easily customizable by other MRI reconstruction experts.

\vspace{-3mm}
\section{Experimental Setup}\label{sec:experimental_setup}
With institutional review board approval and informed consent/assent, pediatric patients presenting with indications of knee pain were scanned using the T2 Shuffling
sequence on a 3 Tesla MRI scanner (GE Healthcare, Waukesha) at Lucile Packard Children's Hospital over a one month period.
A representative case from the scans was used to carry out the design and analysis.

\subsection{Scan Sequence and Reconstruction Parameters}
The acquisition and reconstruction dimensions and sizes are shown in Table~\ref{tab:params}.

\paragraph*{Acquisition}
The T2 Shuffling sequence modified a 3D-FSE acquisition to randomly shuffle and re-acquire phase encode ($y$ and $z$) measurements
during the echo trains.
Scans were performed sagittally at 0.6 mm isotropic resolution with a 16-channel GEM-Flex receive coil array.
The pulse repetition time and echo spacing
were 1200 ms and 6 ms, respectively, and the echo train length was 83. The data from the first three echoes covered the central
portion of k-space and were used to perform ESPIRiT calibration \cite{bib:t2shclinical}.
The remaining $T=80$ echoes were used in the iterative
reconstruction. Each TE was under-sampled with a variable density Poisson Disc distribution, at an acceleration
factor of 139 per time point. Collectively, this represented a relative acceleration of 1.7 and apparent acceleration of 6.9, as defined in Table 1 of \cite{bib:t2shmethod}.
A partial Fourier acceleration of 0.65 was used in the slice direction \cite{bib:t2shpf}.
The acquisition 3D array was $288\times 260\times 240$, and the 
total scan time was 7 minutes and 2 seconds.

\paragraph*{Reconstruction}
The distributed reconstruction was controlled through a Python script, using \texttt{mpi4py} for distributed communication. Each underlying
component of the reconstruction was implemented using BART.
In the reconstruction, $M=2$ soft-SENSE maps were used and $K=4$ subspace coefficient images were reconstructed. The data were
coil compressed from $C_\mathrm{total}=16$ channels to $C=7$ virtual channels.
The fully sampled readout direction ($N_x$) was inverse Fourier transformed, and each slice was independently solved with
$N_\mathrm{itr}=250$ iterations of FISTA. The data were normalized based on the maximum signal value of the raw data and the LLR
regularization parameter was held fixed at $\lambda = 0.0125$ across all scans.
Homodyne filtering was applied to each reconstructed virtual coil image and coil-combined with root sum-of-squares. Gradient
non-linearity correction and DICOM generation was performed using the Orchestra reconstruction library (GE Healthcare) and
the Ox-BART post-processing code, available at \url{http://github.com/mrirecon/ox-bart}.


\subsection{Experimental Compute Cluster}
Each compute node has two (dual-socket) Intel$^{\mbox{\tiny\textregistered}}$ Xeon$^{\mbox{\tiny\textregistered}}$\footnote{Intel and Xeon are trademarks of Intel Corporation or its subsidiaries in the U.S. and/or other countries. Other names and brands may be claimed as the property of others.} Platinum 8180 processors 2.5 GHz. with 28 cores and 38.5 MB total L3 cache per socket. Each node contains 192 GB DDR4 RAM, arranged as 12 DIMMs of 16 GB each, in order to fully utilize the system's available memory channels. The nodes are fully-connected dual-rail (one per socket) Intel$^{\mbox{\tiny\textregistered}}$ Omni-Path Architecture (OPA) 100-gigabit switch. 

Our code is compiled in two parts. Part 1, the BART portion, is compiled with the GNU compiler (gcc 4.8.5), uses FFTW version 3.3.6-pl2 compiled with AVX2 support for optimized transforms, and the Intel$^{\mbox{\tiny\textregistered}}$ Math Kernel Library (MKL) version 18.0.0 for optimized linear algebra routines. Part 2, our optimized kernels described in Section \ref{sec:optimization}, are compiled with the Intel$^{\mbox{\tiny\textregistered}}$ compiler (\texttt{icc}) with the following flags \textit{-O3 -ipo -no-prec-div -fp-model fast=2 -xCORE-AVX512 -mkl -qopenmp -std=c99 -vec-threshold0 -static}. We use MKL for optimized FFT and optimized linear algebra routines. These two parts are linked together with icc. 

We run experiments using \texttt{/dev/shm} as temporary storage space for data between multiple invocations of the BART library. For the timing experiments, we pre-load the input data into \texttt{/dev/shm} and do not count this transfer time in the runtime results. Similarly, we write the output DICOM files to \texttt{/dev/shm} instead of the NFS in order to remove the consideration of network transfer time from our experiments as we observed NFS transfer times to vary widely depending on other jobs running in the cluster.


\begin{table}[!t]
\caption{Runtime breakdown comparison for T2 Shuffling (A) Baseline, (B) Optimizations only, (C) Optimizations with improved quality, and (D) Optimizations with improved quality in the 4-node cluster. Time shown in seconds.}
\label{tab:comparison}
\centering
\small
\begin{tabular}{l r r r r}
  \toprule
  &(A) & (B) & (C) & (D) \\
  Pre-processing & & & & \\
  \midrule
  Conversion to BART Format & 2 & 1 & 1 & 1\\
  Noise Whitening & - & - & 2 & 1\\
  Re-sort and Average & 8 & 4 & 4 & 4 \\
  Coil Compression & 7 & 4 & 4 & 4\\
  Scale Factor & 4 & 0 & - & - \\
  ESPIRiT Maps & 6 & 6 & 6 & 6 \\
  Create Basis & 38 & - & - & -\\
  Re-sort and Project Data & 4 & 1 & 1 & 1 \\
  Broadcast & - & 0 & 0 & 1 \\
  \midrule
  \midrule
  PICS Reconstruction & & & &\\
  \midrule
  Linear Recon. & 20 & -  & - & - \\
  Linear Recon. Process DICOM & 23 & - & - & -\\
  Slice and Detailed Recon. & 271 & - & - & -\\
  Slice and ESPIRiT Maps & - & 13 & 24 & 9\\
  No Slice and Detailed Recon. & - & 94 & 106 & 28\\
  Join Reconstructed Slices & - & 2  & 6 & 1\\
  \midrule
  \midrule
  Post-processing  & & & &\\ 
  \midrule
  Allgather & - & 0 & 0 & 3 \\   
  Process All TEs & 70 & 19 & 18 & 9 \\  
  \midrule
    \midrule
  Other & 6 & 6 & 7 & 15\\
  \textbf{Total} &\textbf{458}  & \textbf{150} & \textbf{180} & \textbf{85}\\
  \bottomrule
\end{tabular}
\squeezeup
\end{table}

\begin{figure}
\centering
\includegraphics[width=0.6\columnwidth]{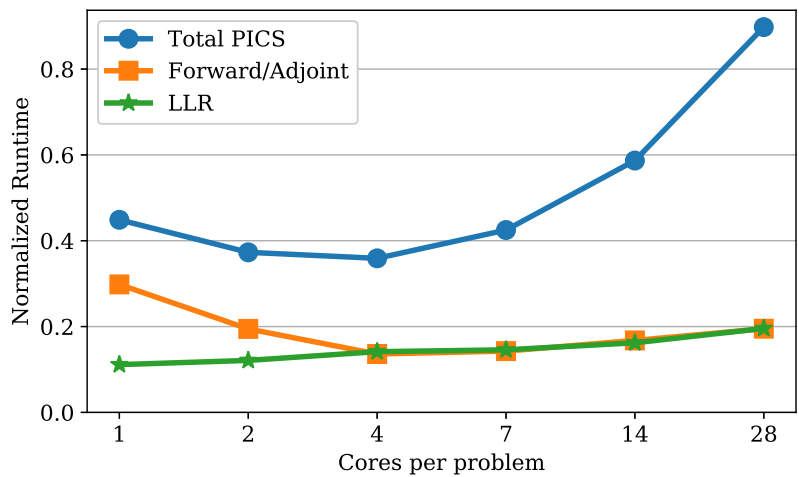}
\caption{PICS runtime for a batch of 288 total slices, varying \#cores assigned per slice. All cores are utilized at each point. Improved performance from 1 core per problem to 4 is due to reduced working set size. Runtime is normalized by number of slices, machines, and iterations.}
\label{fig:cores}
\squeezeup
\end{figure}

\vspace{-2mm}
\subsection{Hospital Compute Cluster}
A four-node system was deployed locally at the hospital, with each node matching the specifications and compilations used
in the experimental compute cluster. The compute cluster is located in a room adjacent to the scanner and connected to the hospital network and MRI scanners with a 1 Gigabit network link. The cluster itself used a 10 Gigabit switch in place of the OPA switch for communication between compute nodes during reconstruction. 

\section{Results}
\label{sec:results}

\subsection{Single-machine Optimization}

In Table~\ref{tab:comparison}, we quantify\footnote{Software and workloads used in performance tests may have been optimized for performance only on Intel microprocessors. 
Performance tests, such as SYSmark and MobileMark, are measured using specific computer systems, components, software, operations and functions. Any change to any of those factors may cause the results to vary. You should consult other information and performance tests to assist you in fully evaluating your contemplated purchases, including the performance of that product when combined with other products.   For more complete information visit www.intel.com/benchmarks.}\textsuperscript{,}\footnote{Performance results are based on testing as of March 31st, 2018, and may not reflect all publicly available security updates. See configuration disclosure for details. No product can be absolutely secure.} the benefits of applying the optimizations described in Section~\ref{sec:optimization},
averaged over five identical reconstructions.
The main improvements to the pre-processing step are a result of the
changes to FFTmod and to parallelizing and caching the basis. Accounting for data slicing and joining,
the detailed PICS reconstruction is sped up from 275 seconds to 109 seconds on one machine.
The post-processing is reduced from 70 to 18 seconds, benefiting from parallel processing.
Overall, the optimizations led to a 3x speedup.
Due to the processing gain, we were able to add reconstruction quality improvements at an additional cost of 30 seconds
on a single machine.

Figure \ref{fig:cores} shows the effect of increasing the number of cores per problem while running many independent PICS slice reconstructions in parallel. All cores on the machine are fully occupied for each point on the graph. In the case of one core per problem, there are 56 independent problems being solved simultaneously. For two cores per problem there are 28 independent simultaneous problems, and so on. As we increase the number of cores per problem, it becomes more likely that the working set derived from processing the $240 \times 260$ slices fits in L2 cache, which is 1 MB/core for our configuration.
The Forward/Adjoint operation per-iteration runtime, LLR per-iteration runtime, and the full solver per-iteration runtime is shown. The Forward/Adjoint runtime and full solver runtimes improve performance with increasing cores up to four due to beneficial caching effects from decreasing the number of simultaneous problems running on the system. Beyond four cores, the overall throughput decreases due to increasing overheads of parallelism as well as the impact of serial portions elsewhere in the solver. As a result, we use four cores per problem in the optimized implementation for the remainder of the paper. 

\subsection{Multi-machine Scaling}

\begin{figure}
\centering
\includegraphics[trim=0 5 10 7, clip, width=0.6\columnwidth]{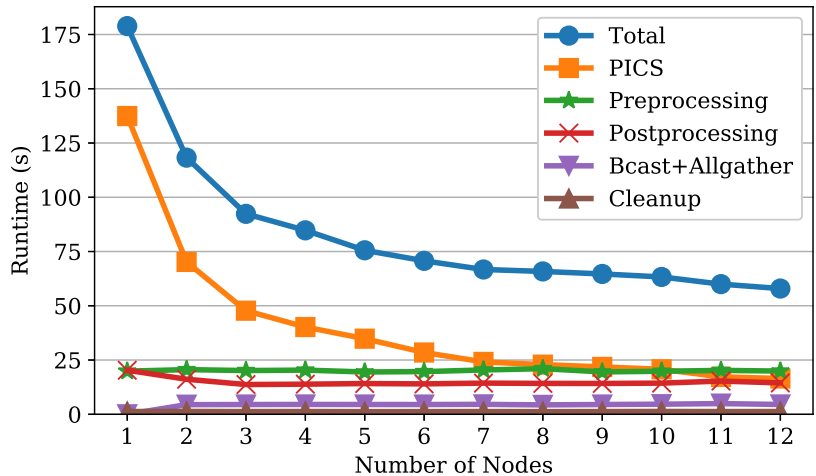}
\caption{Runtime of a T2 Shuffling reconstruction on configurations from 1 to 12 machines. We achieve 60-second total runtime at 11 machines.}
\label{fig:scaling}
\squeezeup
\end{figure}

Figure \ref{fig:scaling} shows the runtime of T2 Shuffling from one to 12 machines. The PICS portion is the only part that is distributed, so therefore this portion scales with the number of machines added. However, the computation is currently limited to one machine for pre-processing and three machines for post-processing so these components quickly become the bottleneck as we increase the number of machines. We didn't choose to distribute the pre and post-processing further so as to hide the distributed computing details from future users of the software who are more productive in the shared-memory BART environment. The four-node configuration, which we deployed in the hospital, provides a total compute time of roughly 85 seconds. 

Since the collectives are implemented in Python using \texttt{mpi4py}, we expect that much of the runtime is being spent loading data into Python from \texttt{/dev/shm} and preparing it for the collective operation, not in the communication itself. This is consistent with the experimental results, in which we see the total time spent in the distributed collective operations Broadcast and Allgather stays roughly constant as we increase the number of machines from two to 12. 

We also see an improvement in the post-processing time from one to three nodes, as we distribute the post-processing to three machines. There is also a sharp improvement in PICS runtime between 10 and 11 nodes. This is the result of load balancing. Since we are doing 14 slices at a time per machine and 288 total slices, 10 nodes gives us exactly 140 slices executed at once on the cluster which completes in two steps, followed by one extra step with one slices on 8 of the 10 machines. At 11 nodes, there are no slices left over after two steps, so the performance improves sharply due to the improved overall load balance. 

\subsection{Hospital Cluster Statistics}
\begin{figure}
  \centering
  \includegraphics[width=0.6\columnwidth]{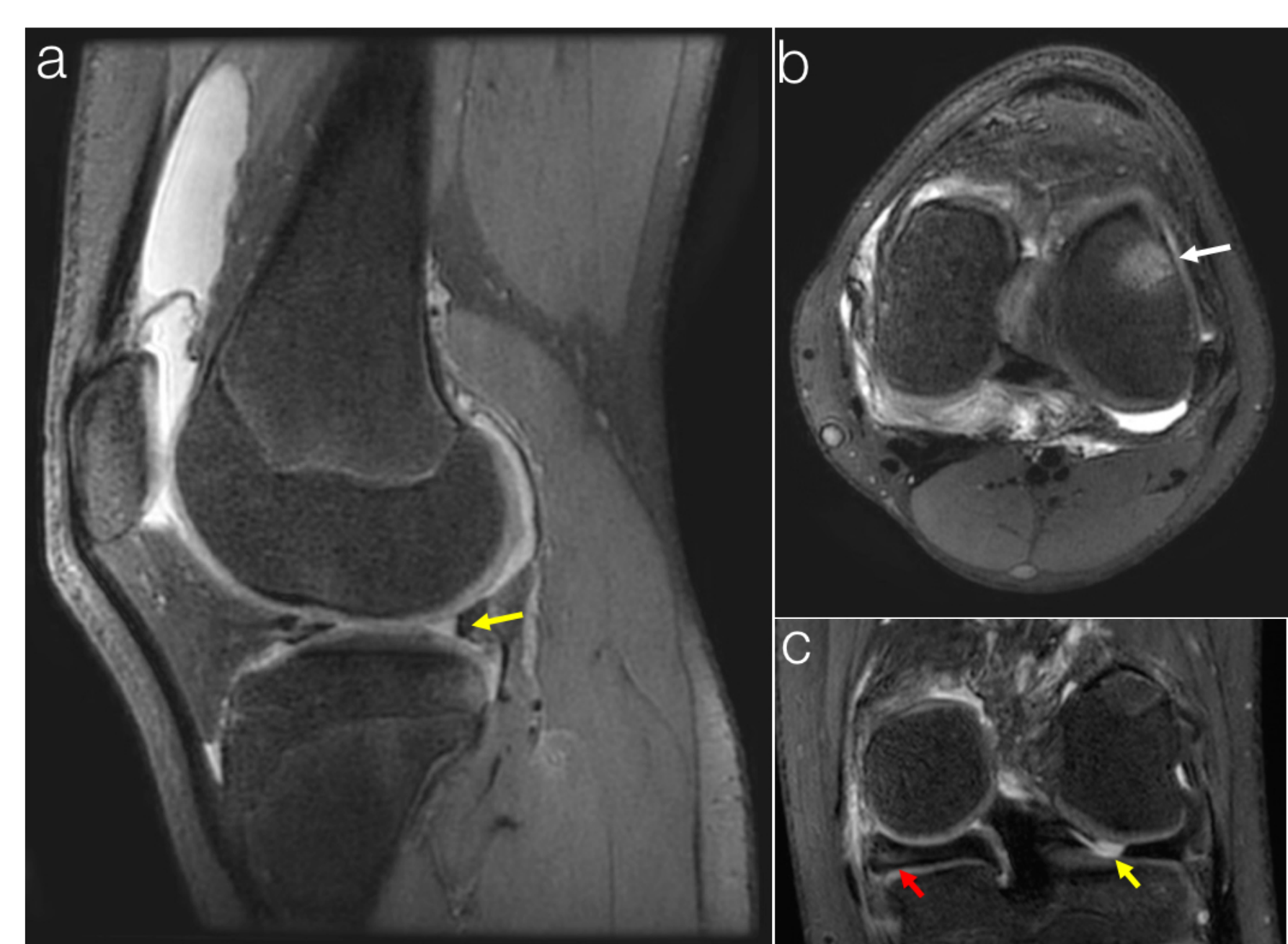}
  \caption{T2 Shuffling reconstruction of pediatric patient presenting with indications of knee pain. The images show
    a torn meniscus (yellow arrows), discoid meniscus (red arrow), and bone marrow edema (white arrow) at effective echo times
  of (a) 21 ms (sagittal reformat), (b) 50 ms (axial reformat), and (c) 90 ms (coronal reformat).}
  \label{fig:knee_image}
\squeezeup
\end{figure}

Over the one-month deployment period, 35 pediatric knee T2 Shuffling scans were performed. The mean, standard deviation, and maximum times for
the main steps of the processing pipeline across these scans are shown in Table~\ref{tab:clinical_runtimes}.
The increase in runtime relative to that reported in Table~\ref{tab:comparison} is attributed to latency associated with reading the raw data and writing the
DICOM images across the 1 Gigabit network. Using 10 Gigabit, as opposed to 100 Gigabit Omnipath for inter-node communication, added about one second to both the Broadcast and the Allgather, which is a small portion of the overall runtime. There is an additional overhead of 10-20 seconds while writing to PACS, which is common to all offline reconstruction implementations.
\begin{table}
  \caption{Reconstruction runtimes for 35 pediatric knee scans over one month period.}
\label{tab:clinical_runtimes}
\centering
\small
\begin{tabular}{l c c c}
  \toprule
  & Mean (s) &  Std. Deviation (s) & Max (s)\\ 
  \midrule
  Pre-processing & 27.97 & 2.70 & 34.40\\
  Broadcast& 2.26 & 0.16 & 2.73\\
  Reconstruction  & 37.24 & 4.14 & 52.60\\
  Allgather & 4.47 & 0.31 & 5.16\\
  Post-processing & 18.27 & 1.59&23.19\\
  Cleanup & 1.13 & 0.07& 1.34 \\
  \midrule
  Total & 91.85 & 7.02 & 108.68\\
  \bottomrule
\end{tabular}
\squeezeup
\end{table}
Figure~\ref{fig:knee_image} shows a representative reconstruction from a pediatric knee patient scanned during the one-month study.
The reconstruction was retrospectively reformatted into 2.5 mm slices in
the sagittal, coronal, and axial planes \cite{bib:t2shclinical} for viewing on the PACS.

\subsection{Multi-Scanner Simulation}
Our current hospital cluster deployment serves four scanners with the 4-node cluster, and is used for T2 Shuffling knee exams.
Here we consider the possible tradeoffs involved with serving many exams from many scanners using compute clusters, if we were to use our current distributed system and reconstruction method for all exams from these scanners. This serves as an upper bound on the amount of computation over the course of the day.

\begin{figure}
\centering
\includegraphics[trim=0 8 5 0, clip, width=0.6\linewidth]{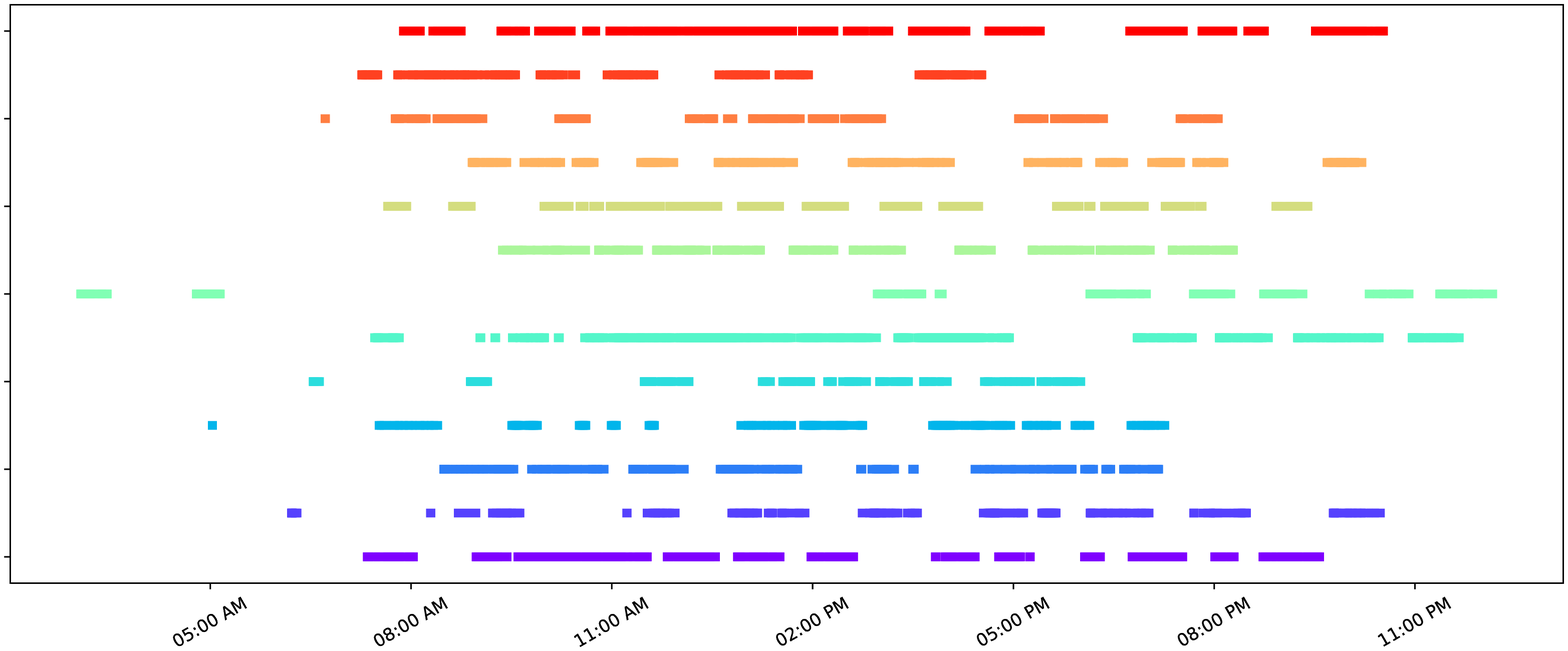}
\caption{MRI scanner schedule from one day out of the one-month study. Each row
  represents an MRI scanner, and each entry shows the duration of a scan. The dotted box shows a zoomed-in portion from the schedule to better visualize the individual scans.}
\label{fig:schedule}
\squeezeup
\end{figure}

Figure~\ref{fig:schedule} shows the schedule from one day out of the one-month study at our institution. Each row represents an MRI scanner, and each entry
shows the duration of a scan. The dotted box shows a zoomed-in portion from the schedule to better visualize the individual scans. With the exception of one
scanner, there is high utilization across the scanners.

Using this data, we can simulate the case where one or more compute clusters are reconstructing images from the scanners. To do so, we first use the image sizes to estimate the runtimes if they were all to employ PICS-based acquisitions and reconstructions. We ignore the issues with transferring data to the clusters and only look at cluster computation time. We also estimate the cost of distributed primitives in the cluster by scaling the measured values from our experiments by the total size of the volume compared to our experimental data. For simplicity, we assumed
that the pre-processing and post-processing times scaled as $O\left(N_xN_yN_z\log(N_xN_yN_z)\right)$, and that the PICS time scaled as $O\left(N_xN_yN_z\log(N_yN_z)\right)$;
that is, the processing is dominated by the FFT factors and linearly scales with the number of concurrent slices. We fit the constants for each stage to match the processing time for the T2 Shuffling knee dataset (Table~\ref{tab:comparison}).

We used the \texttt{simpy} discrete event simulation library to model contention on the cluster. Our model has one process representing the hospital, which creates a new reconstruction process for each scan reconstruction task as they arrive. Each scan reconstruction task tries to access a priority resource representing the clusters. Earlier scans are given higher priority. Once a cluster resource is acquired, the reconstruction process holds the resource for a number of time units equal to our runtime estimates for that particular scan. The time spent waiting and reconstructing are logged for each task throughout the simulated day. 

\begin{figure}
\centering
\includegraphics[trim=0 10 0 0, clip, width=0.6\columnwidth]{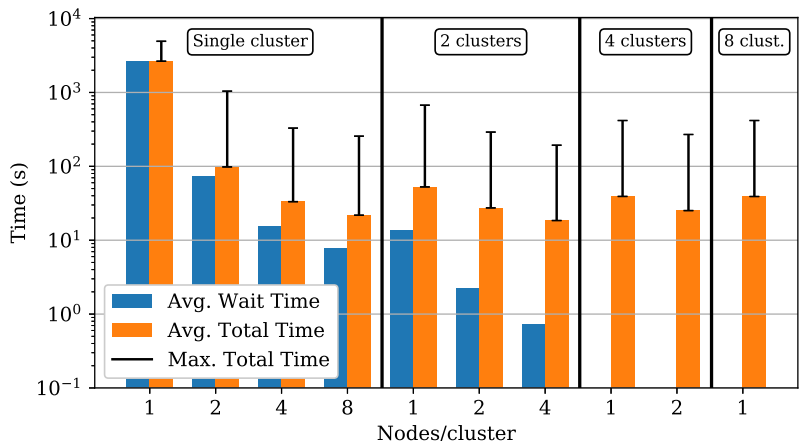}
\caption{Simulated average total reconstruction time and average wait time running reconstruction tasks for all scanners in a large hospital over the course of one day, for different cluster configurations}
\label{fig:simulation}
\squeezeup
\end{figure}

Figure \ref{fig:simulation} shows the result of the simulation. The average wait time is the average time that each reconstruction task spends waiting for the cluster resource. Average total time is the average compute time plus the wait time. The average wait time is reduced significantly when using two clusters rather than one, but there are diminishing returns to adding more clusters after this. We can also examine how to choose the best cluster configuration for a fixed number of machines. Assuming we have eight machines, it makes the most sense, in terms of average and max total reconstruction time, to configure these as two clusters with four nodes per cluster. For this configuration, we can achieve clinically feasible average times (<2min), average wait times near zero, with slightly higher max time of roughly 200 seconds due to a few scans with large reconstruction times.

\squeezeup
\section{Conclusions}\label{sec:conclusions}
Developments to PICS-based acquisitions have shown great promise at reducing MRI scan times, but the reconstruction times have remained a barrier to clinical adoption.
In this work we developed a distributed reconstruction architecture for T2 Shuffling and achieved latency times that make its clinical use feasible.
We have already deployed the system at Lucile Packard Children's Hospital for pediatric knee imaging with great initial success.
Our approach is well-suited for other PICS acquisitions, in which the slices can be independently reconstructed, and it presents an avenue for other
reconstruction formulations to benefit from the distributed optimizations.

Our optimization of the pipeline led to an overall improvement in performance by $5.3 \times$ after speedups and adding additional operations to improve image quality. A $3\times$ speedup came from single-node improvements and a $2.1\times$ speedup came from distributed parallelization on a 4-node cluster. By enabling clinically feasible reconstruction times, a larger number of clinical protocols could potentially leverage PICS acquisitions, leading to shorter overall exam times.
Thus, we expect better scanner utilization as more patients can be accommodated into the same schedule. Dedicated compute clusters could support
this increased utilization without the need for transferring data outside the hospital network.
\squeezeup






\bibliographystyle{IEEEtran}
\bibliography{IEEEfull,bibliography}


\end{document}